\documentclass[a4paper]{article}

\usepackage{amsfonts,amssymb,amsthm,epsfig}

\usepackage[totalwidth=17cm,totalheight=24cm]{geometry}

\begin{document} 
\title{Energy of 4-Dimensional Black Hole}

\author{Dmitriy Palatnik
\thanks{The Waterford, 7445 N Sheridan Rd, Chicago, IL 60626-1818 dmitriy.palatnik@yahoo.com } 
}

\maketitle

\begin{abstract}

In this letter I suggest possible redefinition of mass density or stress-energy for a particle. 
Introducing timelike Killing vector and using Einstein identity $E=mc^2$, I define naturally density of mass 
for general case and calculate energy of black hole.

\end{abstract}

\maketitle

\section{Introduction}

Standard definition of density \cite{Landau}, \cite{Wald}, \cite{Weinberg}, $\mu$, for mass,  
\begin{equation}\label{1}
m = \int dV \sqrt{\gamma} \mu\;,  
\end{equation}
where $\gamma$ is determinant of metric for spacial line element, lacks manifest dependence on speed of motion, $v$, or 
on gravitational interaction with other masses. Since this dependence is important in what follows below, I 
should 'redefine' the mass density. Consider, first, flat spacetime with Cartesian coordinates. 
Matter distribution is specified by $\{ \mu ; U^a \}$, field of density and field of 4-velocity, 
respectively. According to Einstein's formula, 
\begin{equation}\label{0}
m = m_*\left(1-{{v^2}\over{c^2}}\right)^{-{1\over2}}\;,
\end{equation}
{}for the mass field variation,
$\delta m$, moving with speed, $v$, one may write,  
\begin{equation}\label{1.1}
\delta m = \delta m_*\left(1-{{v^2}\over{c^2}}\right)^{-{1\over2}}
 + m_*{{v\delta v}\over{c^2}}\left(1-{{v^2}\over{c^2}}\right)^{-{3\over2}}\;;
\end{equation}
Consider smooth distribution of 3-velocity, $v$, and assume that measuring of density takes place in frame in which 
matter locally at rest, i.e. $v=0$.
{}For differential of mass element measured experimentally by using Newton formula $f=ma$, one obtains,  
\begin{equation}\label{2}
dm = dV {\mu_* \over{\sqrt{1-{v^2\over c^2}}}}\;;
\end{equation}  
here $\mu_*$ is density of mass, measured at rest. Formula (\ref{2}) may be rewritten as
\begin{eqnarray}\label{3} 
dm &=& \mu_* dV{{cdt}\over{ds}} \nonumber \\
   &=& \mu_* {{d\Omega}\over{ds}} \nonumber \\
   &=& \mu_* \sqrt{\gamma} {{d\Omega}\over{ds}} \;;
\end{eqnarray}
where $d\Omega = cdtdV$ is 4-dimensional volume and $ds^2 = g_{ab} dx^adx^b$. In third line of (\ref{3}) I've transfered to
general coordinate system with metric $g_{ab}$. There is an ambiguity in (\ref{3}) because one might put there 
$\sqrt{-g}$ in stead
of $\sqrt{\gamma}$ as well. As it's shown below, $\sqrt{-g}$ seems to be more correct after accepting (with no proof) formula
(\ref{8.2}) for stress-energy; in order to get standard formula (\ref{5}) for stress-energy, I take (\ref{3}) as is. 

By definition (\ref{3}) $\mu_*$ is a (non-scalar) field independent of speed 
in difference with standard density of mass, $\mu$, (\ref{1}), which depends on speed and also not a scalar. Suppose,
now, that mass, $dm$, is at rest in gravitational field $g_{00} = 1 + 2{{\phi(x)}\over{c^2}}$, 
where $\phi(x)$ is gravitational potential. From (\ref{3}) it follows then, 
\begin{eqnarray}\label{4} 
dm &=& \mu_* {{\sqrt{\gamma}}\over{\sqrt{g_{00}}}}{{d\Omega}\over{cdt}} \\
\label{4.1}
   &=& {{\mu_*}\over{\sqrt{g_{00}}}} \sqrt{\gamma} dV \;,
\end{eqnarray}
where from (\ref{1}), (\ref{4.1}) follows $\mu_* = \mu \sqrt{g_{00}}$. 
Using standard technique, i.e. formula 
\begin{equation}\label{4.2} 
\delta S_m = (2c)^{-1}\int d\Omega \sqrt{-g}\,T_{bc}\delta g^{bc}\;,  
\end{equation}
where 
\begin{equation}\label{4.3}
S_m=-mc\int ds\;;
\end{equation}
is action of matter, one obtains stress-energy for
distribution of masses \cite{Landau},
\begin{equation}\label{5}
T^{ab} = {{\mu_* c^2}\over{\sqrt{g_{00}}}} {{dx^a}\over{ds}}{{dx^b}\over{ds}}\;. 
\end{equation}
Connection between standard density of mass and density measured at rest is
\begin{equation}\label{5.1}
\mu = \mu_*{{cdt}\over {ds}}\;.
\end{equation}

Timelike Killing vector field, $\xi_a$, satisfies equations,
\begin{equation}\label{6}
\nabla_b\xi_a + \nabla_a\xi_b = 0 \;;
\end{equation}
suppose, that spacetime is asymptotically flat and $\xi^c\xi_c \rightarrow 1$ on spacial infinity. 
If solution to (\ref{6}) does exist, then, due to stress-energy symmetry $T^{bc}=T^{(bc)}$ and conservation 
$\nabla_c T^{bc}=0$, 
current $J^c=T^{cb}\xi_b$ conserves too: $\nabla_{c}\{ T^{cb}\xi_b\} =0$. Then, due to Gauss theorem 
conserving energy integral does exist, 
\begin{equation}\label{7}
E = \int dV \sqrt{-g} T^{0b}\xi_b\;.
\end{equation}
Accepting Einstein rule, $E=mc^2$, as axiom, one might define mass element according to (\ref{7}), as
\begin{equation}\label{8}
dm = {1\over{c^2}}dE = {1\over{c^2}}dV\sqrt{-g}T^{0c}\xi_c\;.
\end{equation}
{}From (\ref{1}), (\ref{5}), (\ref{8}) it follows formula for standard mass density:
$$\mu = \mu_* {{cdt}\over{ds}}\cdot U^c\xi_c \;,$$
which in general case depends on additional factor $U^c\xi_c$. 
One might attempt to find another action rather than $S_m=-mc\int ds$, because as 
one observes, stress-energy should be taken as 
\begin{equation}\label{8.2}
T^{ab} = {\mu_* c^2} {{dx^a}\over{ds}}{{dx^b}\over{ds}}\;, 
\end{equation}
in stead of (\ref{5}), otherwise using (\ref{5}) in (\ref{7}) one would obtain infinite energy-mass for black hole.
Besides, formulae (\ref{4.2}), (\ref{4.3}), (\ref{8}) lead to inconsistent algebraic equation for stress-energy:
$$T^{ab} = T^{0c}\xi_c{{U^aU^b}\over{U^0}}\;.$$ 
It is most natural to change action (\ref{4.3}), rather than formulae (\ref{4.2}), (\ref{8}). 
For case of Sz metric below, $\xi_cU^c = \sqrt{g_{00}}$.
Note, that in formula (\ref{8.2}) $\mu_*$ is genuine {\em scalar} density of mass; for the
element of mass (\ref{8}) one would obtain,
\begin{equation}\label{8.3}
dm = \mu_*\sqrt{-g}{{d\Omega}\over{ds}}\cdot U^c\xi_c\;.
\end{equation} 
In order to obtain (\ref{8.2}) one should consider actions of type $S_m = -mc\int L_m(\alpha) ds$, where 
$L_m(\alpha)$ is analog of lagrangian, depending only on $\alpha=U^c\xi_c$.

\section{Energy of Sz Black Hole}

Consider Schwarzschild solution,
\begin{equation}\label{9}
ds^2 = \left(1 -
{{2kM}\over{c^2r}}\right)c^2 dt^2 - \left(1 -
{{2kM}\over{c^2r}}\right)^{-1}dr^2 - r^2(d\theta^2 + \sin^2\theta d\phi^2)\;.
\end{equation}
Solving (\ref{6}) for metric (\ref{9}), one obtains timelike Killing vector,
\begin{equation}\label{10}
\xi_a = \left(1 -
{{2kM}\over{c^2r}}, 0, 0, 0\right)\;. 
\end{equation}
Substituting  (\ref{8}), (\ref{8.2}), (\ref{9}), (\ref{10}) in (\ref{7}), and using $dV = drd\theta d\phi$, one gets,
\begin{equation}\label{11}
E = c^2 \int r^2 dr \sin \theta d\theta d\phi \mu_*(r, \theta, \phi)\;.
\end{equation}
Transfering to Cartesian coordinates, $(r, \theta, \phi) \rightarrow (x^1,x^2,x^3)$, and introducing density of mass,
\begin{equation}\label{12}
\mu_* = M \delta (x^1) \delta (x^2) \delta (x^3)\;, 
\end{equation}
one obtains necessary expression for energy of black hole, $E=Mc^2$. Note, that the same result could be derived by
using formula (\ref{8.3}). 

\section{Modification of Killing Equation}

Solutions to Killing eqs (\ref{6}) do exist for restricted class of gravitational configurations; indeed, 4 components
of Killing vector should satisfy 10 equations. Idea of following (i.e. formula (\ref{14}) bellow) belongs to Boris Tsirelson. 
Consider, again, continuous matter distribution, specified by $\{ \mu_*\;; \;U^{c_0}\} $, scalar mass density field,
$\mu_*(x^{c_1})$, and 4-velocity field, $U^{c_0}(x^{c_1})$. Assume, 
that all other than matter fields are absent and only contribution of stress-energy is (\ref{8.2}).  
Due to equations of motion of matter, stress-energy is divergence-free, $\nabla_{c_0}T^{c_0c_1}=0$, and symmetric. 
Then, current $J^{c_0} = T^{c_0c_1}\xi_{c_1}$ does have vanishing divergence, if
\begin{equation}\label{13}
T^{c_0c_1}\left[ \nabla_{c_0}\xi_{c_1}+ \nabla_{c_1}\xi_{c_0}\right] = 0\;.
\end{equation}
{}From conservation of current, $\nabla_{c_0}J^{c_0}=0$, follows conservation of mass-energy (\ref{7}). 
In stead of demanding implementation of Killing eqs (\ref{6}) in order to have (\ref{13}) satisfied, 
consider expression (\ref{8.2}) and demand that Killing vector field satisfies following 4 equations:
\begin{equation}\label{14}
U^{c_1}\left[ \nabla_{c_0}\xi_{c_1}+\nabla_{c_1}\xi_{c_0}\right] =0\;.
\end{equation}
For Sz metric (\ref{9}), solution to (\ref{14}) with 
\begin{equation}\label{U}
U^{c_0}= \left[ (g_{00})^{-{1\over2}},0,0,0\right] \;, 
\end{equation}
is (\ref{10}). Is it true that solving eqs (\ref{14}) for empty space one may consider limit $\mu_* \rightarrow 0$ with
(\ref{U})? One could use (\ref{8.3}) for computing mass-energy of black hole; result is same as above.
One might find solution to (\ref{14}) for Kerr black hole with metric \cite{Landau},
\begin{eqnarray}\label{14-1}
ds^2 &=& \left(1 - {{r_gr}\over{\rho^2}}\right)dt^2 + {{2r_gra}\over{\rho^2}}\sin^2\theta dtd\phi- 
{{\rho^2}\over\Delta}dr^2 - \rho^2d\theta^2\nonumber\\
&& - \left(r^2 + a^2 + {{r_gra^2}\over{\rho^2}}\sin^2\theta\right)\sin^2\theta d\phi^2\;;\\ \label{14-2}
g^{ab}\partial_a\partial_b &=& {1\over\Delta}\left(r^2 + a^2 + {{r_gra^2}\over{\rho^2}}\sin^2\theta\right)
(\partial_t)^2  + {{2r_gra}\over{\rho^2\Delta}}\partial_t\partial_{\phi} - {\Delta\over{\rho^2}}(\partial_r)^2
- {1\over{\rho^2}}(\partial_{\theta})^2 \nonumber\\
&&- {1\over{\Delta\sin^2\theta}}\left(1 - {{r_gr}\over{\rho^2}}\right)(\partial_{\phi})^2\;, 
\end{eqnarray}
where 
\begin{eqnarray}\label{14-3}
\rho^2 &=& r^2 + a^2\cos^2\theta\;; \\ 
\label{14-4}
\Delta &=& r^2 + a^2 -r_gr \;,
\end{eqnarray}
and using expression for 4-velocity,
\begin{equation}\label{U1}
U^{c}=\left[ {1\over{\sqrt{g_{00}+2\dot{\phi}g_{03}+(\dot{\phi})^2g_{33}}}},0,0, {\dot{\phi}
\over{ \sqrt{g_{00}+2\dot{\phi}g_{03}+(\dot{\phi})^2g_{33}}}} \right] \;;
\end{equation}
here $\dot{\phi}\equiv {{d\phi}\over{cdt}}$. 
That is, if to demand $\xi_1=\xi_2=0$, then eqs (\ref{14}) are equivalent to 
\begin{eqnarray}\label{14-7}
\partial_r\xi_0-2\Gamma_{01}^0\xi_0-2\Gamma_{01}^3\xi_3+\dot\phi\{ \partial_r\xi_3-2\Gamma_{13}^0\xi_0-2\Gamma_{13}^3\xi_3\}
&=& 0\;;\\
\label{14-8}
\partial_{\theta}\xi_0-2\Gamma_{02}^0\xi_0-2\Gamma_{02}^3\xi_3+\dot\phi\{ \partial_{\theta}\xi_3-2\Gamma_{23}^0\xi_0-
2\Gamma_{23}^3\xi_3\} &=& 0\;.
\end{eqnarray}
Set of non-zero Christoffel symbols for metric (\ref{14-1}) is $$\Gamma_{01}^0,\;\Gamma_{02}^0,\;\Gamma_{13}^0,\;
\Gamma_{23}^0,\;\Gamma_{00}^1,\;\Gamma_{03}^1,\;\Gamma_{11}^1,\;\Gamma_{12}^1,\;\Gamma_{22}^1,\;\Gamma_{33}^1,$$ 
$$\Gamma_{00}^2,\;\Gamma_{03}^2,\;\Gamma_{11}^2,\;\Gamma_{12}^2,\;\Gamma_{22}^2,\;\Gamma_{33}^2,\;\Gamma_{01}^3,\;
\Gamma_{02}^3,\;\Gamma_{13}^3,\;\Gamma_{23}^3.$$
\begin{equation}\label{16}
\sqrt{-g}=\rho^2\sin\theta\;.
\end{equation}
The solutions of (\ref{14-7}), (\ref{14-8}) are,\footnote{{ }They are also solutions of original Killing eqs (\ref{6}).} 
\begin{eqnarray}\label{15}
\xi_{(1)c_0}&=& \left( g_{00},0,0,g_{03} \right) \;;\;\xi^{c_0}_{(1)} = \left( 1,0,0,0\right) \;;\\
\label{15-1}
\xi_{(2)c_0}&=& \left( g_{03},0,0,g_{33} \right) \;;\;\xi^{c_0}_{(2)} = \left( 0,0,0,1\right)\;;
\end{eqnarray}
here $$g_{00}=1-{{r_gr}\over {\rho^2}}\;;\; g_{03}= {{r_gra}\over{\rho^2}}\sin^2\theta\;;\;
g_{33}=-\left(r^2 + a^2 + {{r_gra^2}\over{\rho^2}}\sin^2\theta\right)\sin^2\theta$$
For energy of Kerr black hole use eqs (\ref{7}), (\ref{8.2}), (\ref{15}); the result is
\begin{equation}\label{17}
E = c^2\int \rho^2 dr \sin\theta d\theta d\phi \, \mu_*(r;\theta ){{g_{00}+\dot{\phi}g_{03}}
\over{g_{00}+2\dot{\phi}g_{03}+(\dot{\phi})^2g_{33}}} \;.
\end{equation}
Transferring to Cartesian coordinates in case $\dot{\phi}=0$, 
according to transformation of spheroidal coordinates to Cartesian, specified in \cite{Anabalon}, 
$(r, \theta , \phi) \rightarrow (x, y, z)$, where
\begin{equation}\label{17-1}
x=\sqrt{r^2+a^2}\sin\theta \cos\phi\;;\;y=\sqrt{r^2+a^2}\sin\theta \sin\phi\;;\;z=r\cos\theta\;;
\end{equation}
and using (\ref{12}), one recovers anew $E=Mc^2$. 
Using second Killing vector, (\ref{15-1}), one obtains conserving integral,
\begin{equation}\label{18}
Q = c^2 \int \rho^2 dr \sin\theta d\theta d\phi \, \mu_*(r;\theta ){{g_{03}+\dot{\phi}g_{33}}
\over{g_{00}+2\dot{\phi}g_{03}+(\dot{\phi})^2g_{33}}} \;.
\end{equation}

\section{A Simple Theorem About Killing Vectors}

Here I should specify a theorem, claiming that if metric of spacetime doesn't depend on coordinate $x^j$, then
spacetime has respective Killing vector $\xi_c=g_{jc}$; number of Killing vectors for spacetime is equal to
number of coordinates on which metric coefficients don't depend. {\em Proof.}
Rewrite Killing eqs (\ref{6}) in form,
\begin{equation}\label{19}
\partial_a\xi_b + \partial_b\xi_a - 2\Gamma_{ab}^c \xi_c =0\;.
\end{equation}
Substitute $\xi_c = g_{jc}$ for specific $j$. Then eq (\ref{19}) reads 
\begin{equation}\label{20}
\partial_jg_{ab}=0\;.
\end{equation}
In case of eqs (\ref{14}) one would have in stead of (\ref{20}) equation $U^a\partial_jg_{ab}=0$. One might even go step back
and use eqs (\ref{13}) then one would obtain equation $U^aU^b\partial_jg_{ab}=0$ in stead of (\ref{20}). 4-velocity,
$U^a$, should be taken along the trajectory of a particle. 
It's understandable that if metric doesn't depend on coordinate(s) $x^j$, then does exist translational
symmetry of the physical system in direction specified by that (or these) coordinate(s), which means that respective
charges (energy, momenta) do conserve.

\section{Acknowledgement}

I wish to thank Christian Network and Boris Tsirelson professor of Tel-Aviv University.\footnote{{ }See, e.g. \cite{KT}} 
I'm grateful for spirit of support from Catholic Church and Lubavich synagogue F.R.E.E.


\begin{thebibliography}{99} 
 

\bibitem{Landau}  L~D~Landau E~M~Lifshits {\em The Classical Theory of Fields} Pergamon Press 1975

\bibitem{KT} L~A~Khalfin B~S~Tsirelson {\em Foundations of Physics} vol 22 No 7 1992 

\bibitem{Wald} Robert~M~Wald {\em General Relativity}  U of Chicago Press 1984

\bibitem{Weinberg} S~Weinberg {\em Gravitation and Cosmology} John Wiley and Sons Inc NY 1972

\bibitem{Anabalon} A~Anabalon {\em et al.} arXiv: gr-qc/1009.3030v1 

\end{thebibliography}
\end{document}